\documentclass{webofc}
\usepackage[varg]{txfonts}   % Web of Conferences font
\usepackage{hyperref}
\hypersetup{
	colorlinks	=true,
	urlcolor	=blue,
	linkcolor	=blue,
	citecolor	=blue,
	pdftitle	={},
	pdfauthor	={Jia-Ming Xie, Jun-Xu Lu, Li-Sheng Geng},
	pdfsubject	={Two-pole structures demystified: chiral dynamics at work}
	}
 
\begin{document}
\title{Dynamical origin of universal two-pole structures and their light quark mass evolution}

\author{\firstname{Jia-Ming} \lastname{Xie}\inst{1} \and
        \firstname{Jun-Xu} \lastname{Lu}\inst{2} \and
        \firstname{Li-Sheng} \lastname{Geng}\inst{3}\fnsep\thanks{\email{lisheng.geng@buaa.edu.cn}}
             \and
        \firstname{Bing-Song} \lastname{Zou}\inst{4}
        % etc.
}

\institute{
School of Physics, Beihang University, Beijing 102206, China 
\and 
School of Physics, Beihang University, Beijing 102206, China
\and
School of
Physics,  Beihang University, Beijing 102206, China\\
Peng Huanwu Collaborative Center for Research and Education, Beihang University, Beijing 100191, China\\
Beijing Key Laboratory of Advanced Nuclear Materials and Physics, Beihang University, Beijing 102206, China \\
Southern Center for Nuclear-Science Theory (SCNT), Institute of Modern Physics, Chinese Academy of Sciences, Huizhou 516000, China
\and
Peng Huanwu Collaborative Center for Research and Education, Beihang University, Beijing 100191, China\\
CAS Key Laboratory of Theoretical Physics, Institute of Theoretical Physics,
Chinese Academy of Sciences, Beijing 100190, China\\
School of Physical Sciences, University of Chinese Academy of Sciences, Beijing 100049, China\\
School of Physics, Peking University, Beijing 100871, China
          }

\abstract{%
Two-pole structures refer to the fact that two dynamically generated states are located close to each other between two coupled channels and have a mass difference smaller than the sum of their widths. Thus, the two poles overlap in the invariant mass distribution of their decay products, creating the impression that only one state exists. This phenomenon was first noticed for the $\Lambda(1405)$ and the $K_1(1270)$, and then for several other states. This report explicitly shows how the two-pole structures emerge from the underlying universal chiral dynamics describing the coupled-channel interactions between a heavy matter particle and a pseudo-Nambu-Goldstone boson. Furthermore, we predict similar two-pole structures in other systems dictated by chiral symmetry, such as the isospin $1/2$
$\bar{K}\Sigma_c-\pi\Xi'_c$ coupled channel, awaiting experimental discoveries. 
}
\maketitle
\section{Introduction}
\label{intro}
The $\Lambda(1405)$, with quantum numbers $J^P=1/2^-$, $I=0$, and $S=-1$, has remained puzzling because it does not easily fit into the constituent quark picture of $qqq$ baryons~\cite{Capstick:1986ter}. Instead, it has long been predicted to be a $\bar{K}N$ bound state~\cite{Dalitz:1959dn}, which is further corroborated by the chiral unitary approaches that combine SU(3)$_L$$\times$SU(3)$_R$  chiral dynamics and elastic unitarity~\cite{Oller:2000ma,Hyodo:2011ur,Oller:2019opk,Mai:2020ltx}.  An unexpected finding of the chiral unitary approaches is that the $\Lambda(1405)$ corresponds to two dynamically generated poles in the second Riemann sheet of the complex energy plane~\cite{Oller:2000fj,Jido:2003cb,Meissner:2020khl}, between the thresholds of $\pi \Sigma$ and $\bar{K}N$. Such a two-pole picture has been confirmed in numerous leading order~\cite{Jido:2003cb}, next-to-leading order~\cite{Oller:2000fj,Ikeda:2012au,Guo:2012vv,Mai:2012dt}, and even next-to-next-to-leading order studies~\cite{Lu:2022hwm}. In the following years, it was shown that the  $K_1(1270)$~\cite{Roca:2005nm,Geng:2006yb} and $D_0^*(2300)$~\cite{Kolomeitsev:2003ac,Guo:2006fu,Guo:2009ct,Albaladejo:2016lbb,Guo:2018tjx,Du:2020pui}  also correspond to two poles, needed to explain many relevant experimental data~\cite{Geng:2006yb,Du:2020pui} or lattice QCD data~\cite{Asokan:2022usm}.

The fact that such two-pole structures emerge in many different sectors requires an explanation.  Ref.~\cite{Jido:2003cb} shows that one expects three bound states, one singlet and two degenerate octets in the SU(3) flavor symmetry limit.  In the physical world where SU(3) symmetry is broken, the singlet develops into the lower pole of the $\Lambda(1405)$, and one octet evolves into the higher pole.  Similar arguments have been made for $K_1(1270)$~\cite{Roca:2005nm} and $D_0^*(2300)$~\cite{Albaladejo:2016lbb}. We note that the exact flavor contents of the two poles of the $\Lambda(1405)$ need further investigation according to the recent next-to-leading order study~\cite{Guo:2023wes}.

This talk explicitly demonstrates how the two-pole structures emerge from the underlying coupled-channel chiral dynamics and the pseudo-Nambu-Goldstone~(pNG) nature of the pseudoscalar mesons. We note that the present analysis complements the group-theoretical analyses of Refs.~\cite{Jido:2003cb,Roca:2005nm,Albaladejo:2016lbb}, where one expects two bound states in the SU(3) symmetry limit, which evolve into those observed in the physical world. However, such analyses do not explicitly tell the dynamical origin of the two-pole structures and why they show up as one resonant state in the invariant mass distribution of their decay products. In this sense, the $D_0^*(2300)$ differs from the $\Lambda(1405)$ and $K_1(1270)$ because the two poles corresponding to $D_0^*(2300)$ are well separated and cannot be misidentified as a single state. In particular, we would like to answer the following three questions. 1) Does the off-diagonal coupling between the two dominant channels play a decisive role?  2) Is the energy dependence of the Weinberg-Tomozawa potential relevant? 3) How does the explicit breaking of chiral symmetry generate the two-pole structures?

\section{Coupled-channel chiral dynamics}
\label{sec-1}
First, we would like to point out that the coupled-channel chiral dynamics dynamically generating the $\Lambda(1405)$ and $K_1(1270)$ states share the same form, i.e., the so-called Weinberg-Tomozawa term. 
For the $\Lambda(1405)$, the leading order chiral potential in the center of mass  frame reads~\cite{Jido:2003cb}:
\begin{equation}
    V_{ij}=-\frac{C_{ij}}{4f^2}\left(2\sqrt{s}-M_{i}-M_{j}\right)=-\frac{C_{ij}}{4f^2}\left(E_i+E_j\right),
    \label{VPB}
\end{equation}
where the subscripts $i$ and $j$ represent the incoming and outgoing channels in isospin basis, $M$ is the baryon mass, $E$ is the pseudoscalar meson energy, and $C_{ij}$ are the corresponding Clebsch-Gordan~(CG) coefficients. Note that we have neglected the three momentum of the baryon in comparison with its mass.

Likewise for the pseudoscalar-vector interaction, the leading order $S$-wave chiral potential reads~\cite{Birse:1996hd,Roca:2005nm} 
\begin{equation}
    V_{ij}\left(s\right)=-\epsilon^i \cdot \epsilon^j \frac{C_{ij}}{8f^2}\left[ {3s-\left(M_i^2+m_i^2+M_j^2+m_j^2\right)}{-\frac{1}{s}\left(M_i^2-m_i^2\right)\left(M_j^2-m_j^2\right)}\right].
    \label{VPV}
\end{equation}
Note that $M_{i,j}$  are the vector mesons' masses, and $m_{i,j}$ are those of the pseudoscalar mesons. Close to the threshold, considering the light masses of the pseudoscalar mesons as well as the chiral limit of $M_i=M_j\equiv M$, Eq.~(\ref{VPV}) reduces to
\begin{equation}
     V_{ij}\left(s\right)=-\epsilon^i \cdot \epsilon^j \frac{C_{ij}}{8f^2}4M\left(E_i+E_j\right),
\end{equation}
which is the same as Eq.~(\ref{VPB}) up to the scalar product of polarization vectors, trivial dimensional factors, and CG coefficients. 

Comparing the two potentials, we can conclude that the two-pole structures of $\Lambda(1405)$ and $K_1(1270)$ have the same origin. Furthermore, for the $\Lambda(1405)$, the most relevant channels are $\bar{K}N$ and $\pi\Sigma$, while   for the $K_1(1270)$ the most relevant channels are $\rho K$ and $K^*\pi$. As a result, in this work, we omit other less relevant channels and adopt the two-channel approximation, and, in addition, we focus on the case of $\Lambda(1405)$. The same discussions apply to the $K_1(1270)$.

\subsection{Off-diagonal coupling}
\label{sec-2}
One may naively expect that the two poles are linked to the coupling between the $\bar{K}N$ and $\pi\Sigma$ channels. 
This is not the case. We can conclude by multiplying a factor $0\le x\le 1$ to the off-diagonal matrix elements of the Weinberg-Tomozawa potential and obtaining the evolution of the two poles that: 1) Even in the limit
of complete decoupling, \textit{i.e.}, the so-called zero coupling limit ($x = 0$)~\cite{Hyodo:2007jq,Cieply:2016jby}, the two poles still appear in between
the $\Bar{K}N$ and $\pi\Sigma$ thresholds. 2) the coupling between the two channels not only pushes the two poles higher but also allows the higher pole to decay into the $\pi\Sigma$ channel and, as a result, develops a finite width. Nevertheless, the most important issue is that the coupling between the two channels is not the driving factor for two dynamically generated states between the two relevant channels.  On the other hand, it does play a role in the emergence of the two-pole structure because otherwise, the higher pole will not manifest itself in the invariant mass distribution of the lower channel.

\subsection{Energy dependence of the Weinberg-Tomozawa term}

It is easy to see that by turning off the chiral potential's energy dependence, one cannot generate two states between the $\bar{K}N$ and $\pi\Sigma$ channels. Instead, one can only obtain one bound state and one virtual state or two bound states by fine-tuning the strength of the chiral potential. 

 \subsection{Explicit chiral symmetry breaking}
 The following demonstrates how the underlying chiral dynamics generate the two-pole structure of the $\Lambda(1405)$.  As shown above, the coupling between $\bar{K}N$ and $\pi\Sigma$ is not decisive in developing the two-pole structure. Therefore, we focus on the single-channel approximation to simplify the discussion.
 According to Eq.~(\ref{VPB}), the diagonal Weinberg-Tomozawa interaction is proportional to the energy of the pseudoscalar meson. Such a form is the only one consistent with chiral symmetry and its breaking and thus of universal nature. For the $\Bar{K}N$ and $\pi\Sigma$ channels of our interest, they read~
\begin{equation}
\begin{aligned}
    V_{\Bar{K}N- \Bar{K}N}\left(\sqrt{s}\right)=&-\frac{6}{4f^2}E_{\Bar{K}}=-\frac{6}{4f^2}\sqrt{m_{\Bar{K}}^2+q_{\Bar{K}}^2},\\
    V_{\pi \Sigma- \pi \Sigma}\left(\sqrt{s}\right)=&-\frac{8}{4f^2}E_{\pi}=-\frac{8}{4f^2}\sqrt{m_\pi^2+q_{\pi}^2}.
    \label{V-diagonal}
\end{aligned}
\end{equation}
Due to the explicit breaking of chiral symmetry,  the kaon mass is much larger than the pion mass. As a result, close to their respective thresholds, the $\bar{K}N$ interaction is stronger than the $\pi \Sigma$ one, leading to a $\bar{K}N$ bound state. In addition, the energy dependence and the small pion mass enhance the $q^2$ term of the $\pi\Sigma$ interaction and, therefore, are responsible for the $\pi\Sigma$ resonance. We note in passing that in the chiral limit, the $\bar{K}N$ and $\pi\Sigma$ thresholds are degenerate. The breaking of the degeneracy is of second order in the chiral power counting and, therefore, relatively small, which is also necessary for the emergence of the two-pole structure . 

\begin{figure*}[htpb]
    \centering
    \includegraphics[width=2.4 in]{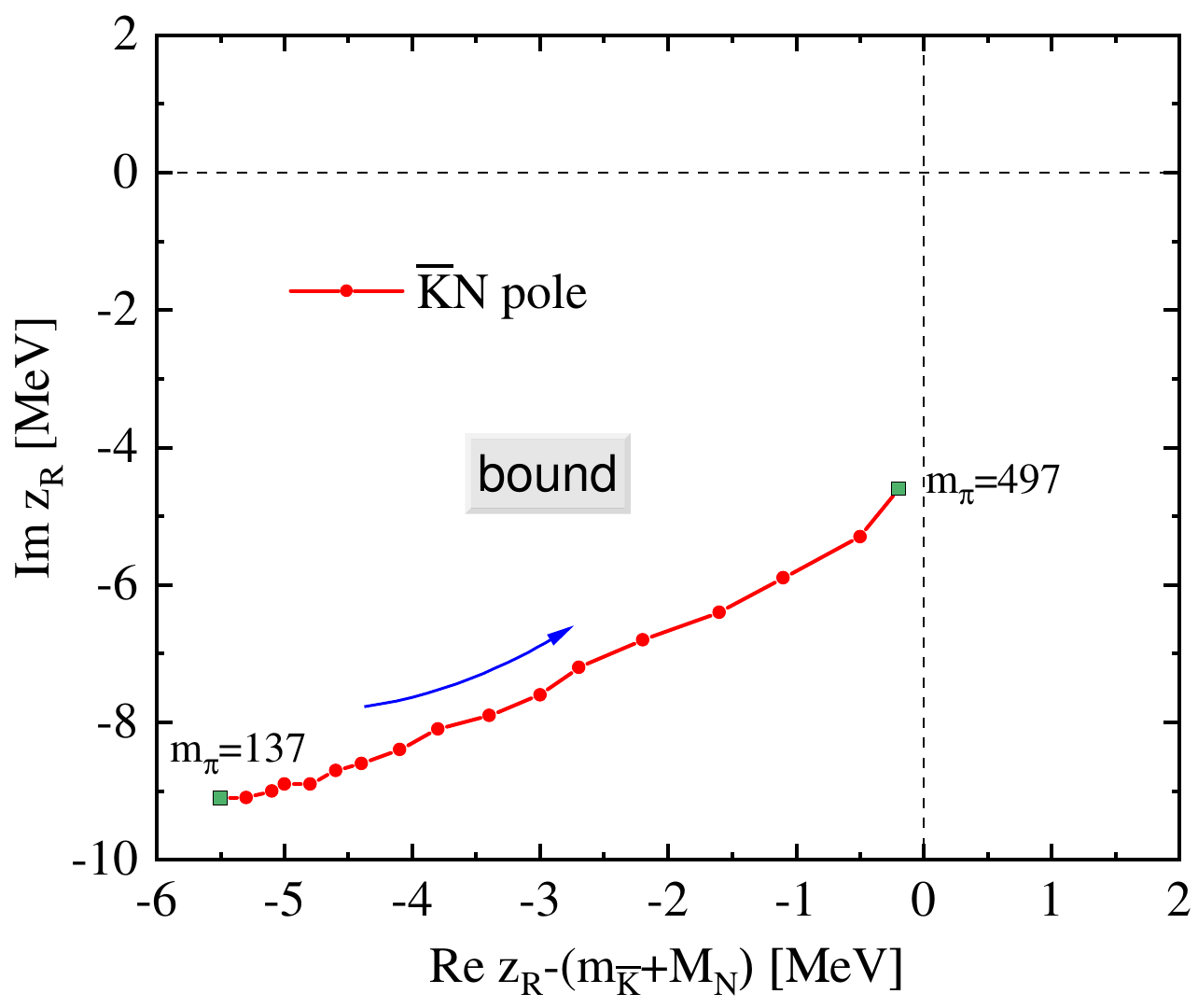}\quad
       \includegraphics[width=2.5 in]{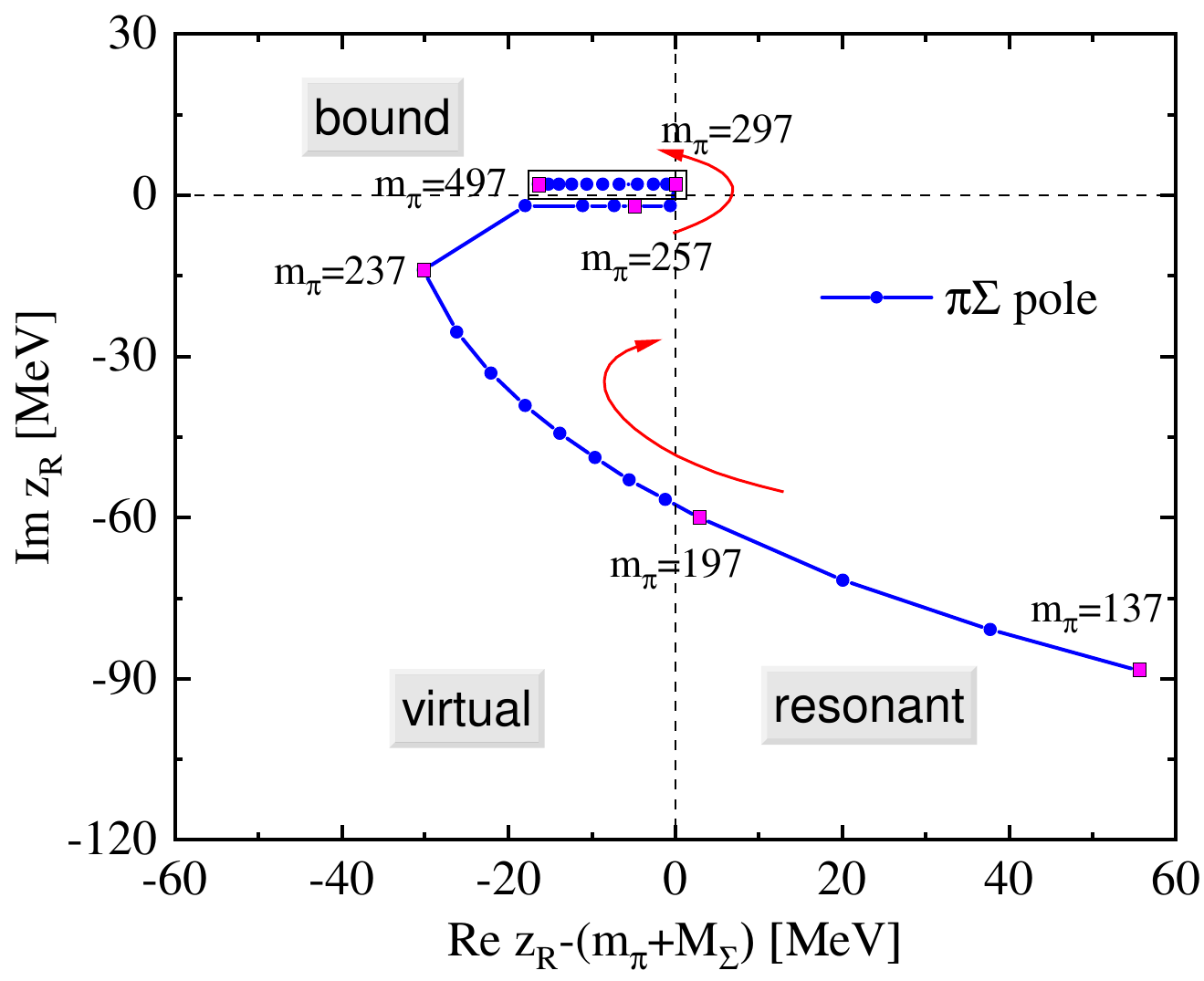}
    \caption{Trajectories of the two poles of $\Lambda(1405)$ as a function of the pion mass  $m_{\pi}$ from 137 MeV to 497 MeV. Critical masses are labeled by solid squares, between which the points are equally spaced.}
    \label{fig:2channel}
\end{figure*}

From the above discussion, it is clear that the dynamical origin of the two-pole structure can be verified by studying the two-pole trajectories as a function of the light-quark (pion) mass. As the pion mass changes, the masses of the baryons and the kaon also vary. We adopt the covariant baryon chiral perturbation theory to describe their light-quark mass dependence and follow the lattice QCD simulations of the PACS-CS Collaboration~\cite{PACS-CS:2008bkb} to determine the light-quark mass dependence of the involved hadrons~\cite{Ren:2012aj}.  
The trajectories of the two poles of $\Lambda(1405)$ are shown in Fig.~\ref{fig:2channel}. The evolution of the higher pole is simple. As the pion mass increases, both its real and imaginary parts decrease. Note that as the pion mass increases, the two thresholds also increase. On the other hand, the trajectory of the lower pole is more complicated and highly nontrivial. As the pion mass increases, it first becomes a virtual state from a resonant state for a pion mass of 200 MeV. For a pion mass of 300 MeV, it becomes a bound state and remains so up to the pion mass of 500 MeV. One should note that the pole trajectories are tied to the quark mass dependence of the involved hadrons and other relevant quantities, such as the decay and subtraction constants. As a result, caution should be taken to directly compare Fig.~\ref{fig:2channel} with future lattice QCD simulations. Instead, Fig.~\ref{fig:2channel} should be viewed qualitatively. 

\subsection{More two-pole structures}
\begin{figure}[htpb]
\centering
    \includegraphics[width=2.4 in]{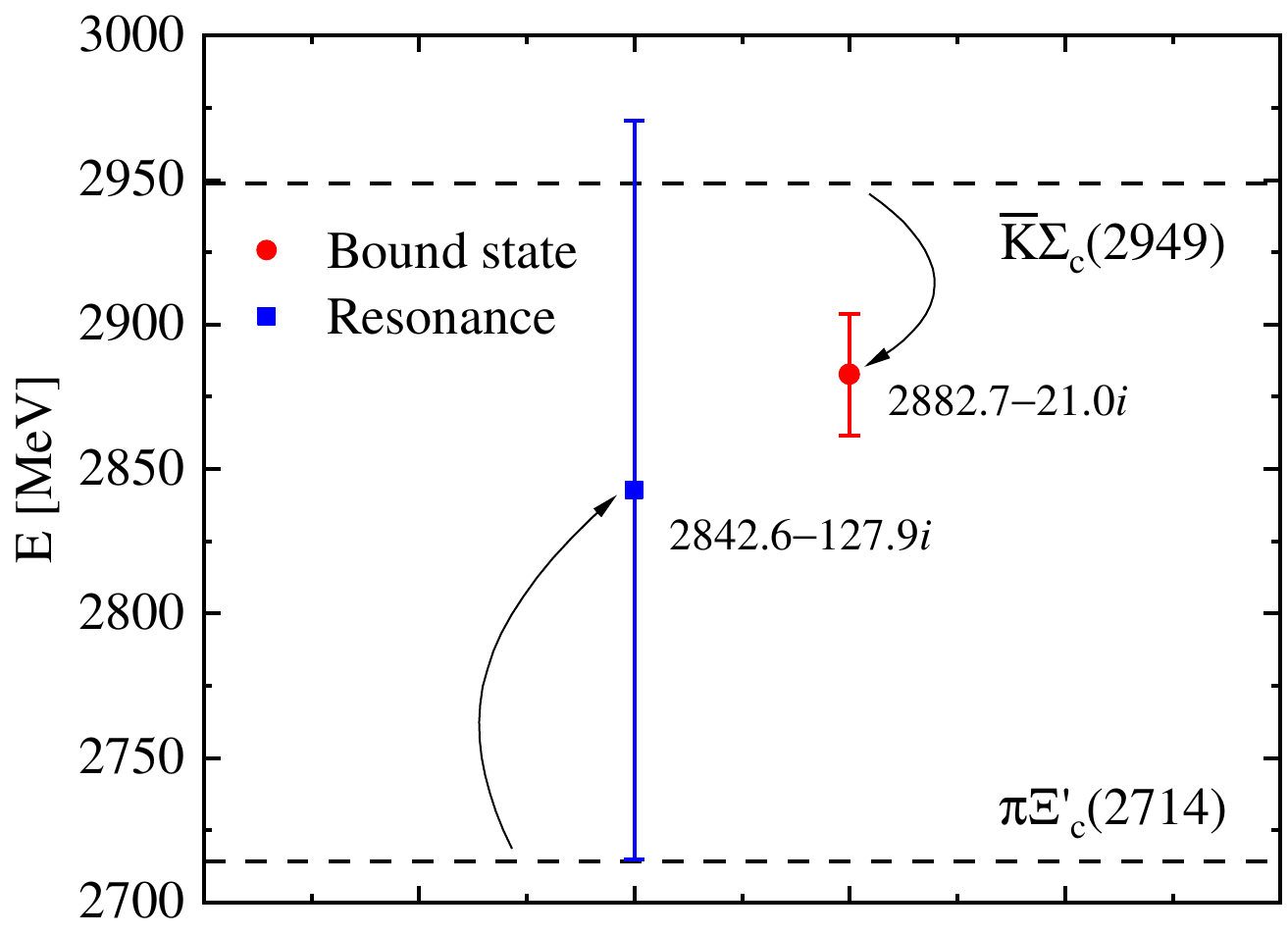}
    \caption{Two poles of the $\Bar{K}\Sigma_c-\pi \Xi_c'$ system. 
    The dominant channels in relation to the two states are denoted by the arrows. The vertical bars are the widths corresponding to twice of the imaginary parts of the pole positions.}
    \label{fig:single charm baryon}
\end{figure}

In principle, because of the universality of chiral dynamics discussed in this work, one can expect more such two-pole structures in other systems by replacing the octet baryon or vector meson with any other heavy matter particle, such as a singly charmed baryon. Using the criteria proposed in this work, one expects that the $\Bar{K}\Sigma_c(2949)$ and $\pi\Xi_c'(2714)$ coupled-channel system can develop a  two-pole structure~\cite{Xie:2023cej} (see Fig.~\ref{fig:single charm baryon}).  
This is also the case for the $\Xi(1820)$ state~\cite{Molina:2023uko}. For the $D_0^*(2300)$ state, because the two poles are well separated, it does not make much sense to talk about a two-pole structure, at least not in the present context. The same can be said for the $f_0(500)$ and $f_0(980)$ states dynamically generated by the $\pi\pi$ and $K\bar{K}$ coupled-channel chiral dynamics. 

\section{Summary and outlook}
Several two-pole structures have been identified and extensively studied in the past two decades. In this talk, complementary to the widely accepted group-theoretical analyses, we demonstrated the dynamical origin of this fascinating phenomenon. We showed explicitly how the universal chiral dynamics, particularly its explicit breaking, is responsible for the emergence of two poles between two coupled channels, one resonance with respect to the lower channel and one bound state with respect to the higher channel. The difference of their masses is smaller than the sum of their widths. In such a way, they can be potentially misidentified as one single state. 

We proposed that one can study the pole trajectories of the two poles of the $\Lambda(1405)$ (and the other two-pole structures) on the lattice to verify the role of explicit chiral symmetry breaking and the dynamical origin of the two poles. The latest results from the Baryon Scattering Collaboration~\cite{BaryonScatteringBaSc:2023ori,BaryonScatteringBaSc:2023zvt} for a single pion mass of about 200 MeV, found two poles, one bound and one virtual, in the $\bar{K}N$-$\pi\Sigma$ system, providing a nontrivial verification of the two-pole structure of the $\Lambda(1405)$ and the idea proposed in Ref.~\cite{Xie:2023cej}. However, it will be more relevant if one can perform simulations for multiple pion masses. Studies of other two-pole structures on the lattice and experimentally are also strongly recommended. 

Finally, we would like to point out that the two-pole structures discussed in this work are of dynamical origin. They should not be confused with those discussed in other contexts, such as those of Ref.~\cite{Zhou:2020moj}, where one is of dynamical origin and the other belongs to the so-called CDD poles~\cite{Castillejo:1955ed}. 

\section{Acknowledgment}This work is partly supported  by the National Key R\&D Program of China under Grant No. 2023YFA1606700, the National Natural Science Foundation of China under Grant  No.11975041,  No.11961141004, No.12105006, No. 11835015, No. 12047503, No. 12070131001, the Deutsche Forschungsgemeinschaft (DFG Project-ID 196253076-TRR 110), and the Grant of Chinese Academy of Sciences (XDB34030000).
\bibliography{menu.bib}

\begin{thebibliography}{34}

\bibitem{Capstick:1986ter}
S.~Capstick, N.~Isgur, Phys. Rev. D \textbf{34}, 2809 (1986)

\bibitem{Dalitz:1959dn}
R.H. Dalitz, S.F. Tuan, Phys. Rev. Lett. \textbf{2}, 425 (1959)

\bibitem{Oller:2000ma}
J.A. Oller, E.~Oset, A.~Ramos, Prog. Part. Nucl. Phys. \textbf{45}, 157 (2000), \texttt{hep-ph/0002193}

\bibitem{Hyodo:2011ur}
T.~Hyodo, D.~Jido, Prog. Part. Nucl. Phys. \textbf{67}, 55 (2012), \texttt{1104.4474}

\bibitem{Oller:2019opk}
J.A. Oller, Prog. Part. Nucl. Phys. \textbf{110}, 103728 (2020), \texttt{1909.00370}

\bibitem{Mai:2020ltx}
M.~Mai, Eur. Phys. J. ST \textbf{230}, 1593 (2021), \texttt{2010.00056}

\bibitem{Oller:2000fj}
J.A. Oller, U.G. Mei{\ss}ner, Phys. Lett. B \textbf{500}, 263 (2001), \texttt{hep-ph/0011146}

\bibitem{Jido:2003cb}
D.~Jido, J.A. Oller, E.~Oset, A.~Ramos, U.G. Meissner, Nucl. Phys. A \textbf{725}, 181 (2003), \texttt{nucl-th/0303062}

\bibitem{Meissner:2020khl}
U.G. Mei\ss{}ner, Symmetry \textbf{12}, 981 (2020), \texttt{2005.06909}

\bibitem{Ikeda:2012au}
Y.~Ikeda, T.~Hyodo, W.~Weise, Nucl. Phys. A \textbf{881}, 98 (2012), \texttt{1201.6549}

\bibitem{Guo:2012vv}
Z.H. Guo, J.A. Oller, Phys. Rev. C \textbf{87}, 035202 (2013), \texttt{1210.3485}

\bibitem{Mai:2012dt}
M.~Mai, U.G. Mei{\ss}ner, Nucl. Phys. A \textbf{900}, 51  (2013), \texttt{1202.2030}

\bibitem{Lu:2022hwm}
J.X. Lu, L.S. Geng, M.~Doering, M.~Mai, Phys. Rev. Lett. \textbf{130}, 071902 (2023), \texttt{2209.02471}

\bibitem{Roca:2005nm}
L.~Roca, E.~Oset, J.~Singh, Phys. Rev. D \textbf{72}, 014002 (2005), \texttt{hep-ph/0503273}

\bibitem{Geng:2006yb}
L.S. Geng, E.~Oset, L.~Roca, J.A. Oller, Phys. Rev. D \textbf{75}, 014017 (2007), \texttt{hep-ph/0610217}

\bibitem{Kolomeitsev:2003ac}
E.E. Kolomeitsev, M.F.M. Lutz, Phys. Lett. B \textbf{582}, 39 (2004), \texttt{hep-ph/0307133}

\bibitem{Guo:2006fu}
F.K. Guo, P.N. Shen, H.C. Chiang, R.G. Ping, B.S. Zou, Phys. Lett. B \textbf{641}, 278 (2006), \texttt{hep-ph/0603072}

\bibitem{Guo:2009ct}
F.K. Guo, C.~Hanhart, U.G. Meissner, Eur. Phys. J. A \textbf{40}, 171 (2009), \texttt{0901.1597}

\bibitem{Albaladejo:2016lbb}
M.~Albaladejo, P.~Fernandez-Soler, F.K. Guo, J.~Nieves, Phys. Lett. B \textbf{767}, 465 (2017), \texttt{1610.06727}

\bibitem{Guo:2018tjx}
Z.H. Guo, L.~Liu, U.G. Mei\ss{}ner, J.A. Oller, A.~Rusetsky, Eur. Phys. J. C \textbf{79}, 13 (2019), \texttt{1811.05585}

\bibitem{Du:2020pui}
M.L. Du, F.K. Guo, C.~Hanhart, B.~Kubis, U.G. Mei\ss{}ner, Phys. Rev. Lett. \textbf{126}, 192001 (2021), \texttt{2012.04599}

\bibitem{Asokan:2022usm}
A.~Asokan, M.N. Tang, F.K. Guo, C.~Hanhart, Y.~Kamiya, U.G. Mei\ss{}ner (2022), \texttt{2212.07856}

\bibitem{Guo:2023wes}
F.K. Guo, Y.~Kamiya, M.~Mai, U.G. Mei\ss{}ner, Phys. Lett. B \textbf{846}, 138264 (2023), \texttt{2308.07658}

\bibitem{Birse:1996hd}
M.C. Birse, Z. Phys. A \textbf{355}, 231 (1996), \texttt{hep-ph/9603251}

\bibitem{Hyodo:2007jq}
T.~Hyodo, W.~Weise, Phys. Rev. C \textbf{77}, 035204 (2008), \texttt{0712.1613}

\bibitem{Cieply:2016jby}
A.~Ciepl\'y, M.~Mai, U.G. Mei\ss{}ner, J.~Smejkal, Nucl. Phys. A \textbf{954}, 17 (2016), \texttt{1603.02531}

\bibitem{PACS-CS:2008bkb}
S.~Aoki et~al. (PACS-CS), Phys. Rev. D \textbf{79}, 034503 (2009), \texttt{0807.1661}

\bibitem{Ren:2012aj}
X.L. Ren, L.S. Geng, J.~Martin~Camalich, J.~Meng, H.~Toki, JHEP \textbf{12}, 073 (2012), \texttt{1209.3641}

\bibitem{Xie:2023cej}
J.M. Xie, J.X. Lu, L.S. Geng, B.S. Zou, Phys. Rev. D \textbf{108}, L111502 (2023), \texttt{2307.11631}

\bibitem{Molina:2023uko}
R.~Molina, W.H. Liang, C.W. Xiao, Z.F. Sun, E.~Oset (2023), \texttt{2309.03618}

\bibitem{BaryonScatteringBaSc:2023ori}
J.~Bulava et~al. (Baryon Scattering (BaSc)) (2023), \texttt{2307.13471}

\bibitem{BaryonScatteringBaSc:2023zvt}
J.~Bulava et~al. (Baryon Scattering (BaSc)) (2023), \texttt{2307.10413}

\bibitem{Zhou:2020moj}
Z.Y. Zhou, Z.~Xiao, Eur. Phys. J. C \textbf{81}, 551 (2021), \texttt{2008.08002}

\bibitem{Castillejo:1955ed}
L.~Castillejo, R.H. Dalitz, F.J. Dyson, Phys. Rev. \textbf{101}, 453 (1956)

\end{thebibliography}

\end{document}